# Extremum Power Seeking Control of A Hybrid Wind-Solar-Storage DC Power System

Dan Shen, Afshin Izadian, *Senior Member, IEEE*
Energy Systems and Power Electronics Laboratory
Purdue School of Engineering and Technology, Indianapolis 46202, USA
aizadian@iupui.edu

*Abstract*—this paper presents a combined power system with a common dc bus which contains solar power, wind power, battery storage and a constant power dc load (CDL). In wind system, an AC-DC uncontrolled rectifier is used at first stage and the DC-DC converter is controlled by a maximum power point tracker (MPPT) at the second stage. In the solar system, two cascaded boost converters are controlled through a sliding mode controller (SMC) to regulate the power flow to a load. A supervisory control strategy is also introduced to maximize the simultaneous energy harvesting from both renewable sources and balance the energy between the sources, battery and the load. According to the level of power generation available at each renewable energy source, the state of charge in the battery, and the load requirement, the controller operation results in four contingencies. Simulation results show accurate operation of the supervisory controller and functionality of the maximum power point tracking algorithm for solar and wind power sources.

## I. Introduction

Recently, there have been concerns on global climate deterioration and its related harmful effect such as environmental pollution and sustainable development problems. As a solution, clean renewable resources have been given increasing interest [1]. Compared with other new energy technologies, wind and solar have been set up as proven future sources of energy because of their environment-friendly, abundant and cost-effective utilization characteristics. Harnessing these two energies for electric power generation is the area of aiming at quality and reliability in the electricity delivery [2], [3], [4]. However, there are some difficulties associated with wind and solar in power system, e.g. intermittency of wind or solar and instability of the load. Accordingly, photovoltaic (PV) arrays, wind turbines and batteries are used to feed a dc or ac bus connected to the load, as well as the utility grid, constituting the so-called micro-grid [5], [6]. Micro-grids operate in both standalone and grid connected modes. The wind and solar sources can compensate each other and their simultaneous intermittency is complemented by the use of an energy storage device [7].

Recent advances in dc distribution systems have shown several advantages with respect to ac systems. First, dc system provides higher power quality with low harmonic content [8], [9]. Secondly, the switch mode power converters and their current limiting features can provide an excellent uninterrupted power handling [10], and various sources of power can be combined into a standalone micro-grid [11], [12]. Overall end-to-end energy conversion efficiency in a DC micro-grid can reach 77%-85% whereas a less than 60% in AC counterparts. Most of this efficiency improvement is because of the elimination of AC loads and their associated power inverters [13]. Most of the designs have employed a dc/dc power converter to interface the battery to enable the charge–discharge process. While, a battery can be directly connected to the dc bus and be regulated by limited variation of DC bus voltage. Therefore, both the wind power and solar power can operate as a voltage source to inject power to the load and battery. Supervisory controller determines the source type according to the availability and rating of the power. As the DC bus voltage floats, the battery can be charged or discharged. This new hybrid system improves the stability and reliability of power supply. Maximum power point tracking is continuously operated to maximize the power from wind and solar [14-19].

In this paper, the Perturbation and Observation (P&O) method is used on the wind power sources and the adaptive sliding mode MPPT control is applied in the PV system. The cascade configuration of two power converters is used for each source. In this case, the first and second stage experience a constant power load (CPL). This destabilizes the operation of the power converters as the rate of voltage and current variations in the system show a negative slope. Sliding mode controller and the storage device can be utilized to stabilize the operation of the power converters in solar and wind power systems respectively.

This paper is organized as follows. The proposed hybrid energy generation system is given in section II. Section III introduces the wind power system and the MPPT algorithm based on P&O method. Section IV describes the Extremum Seeking Control (ESC) MPPT for solar system by utilizing the sliding mode theory through a cascaded boost converters working as LFRs. Supervisory control strategy is discussed in section V. Finally, the conclusions of the paper are summarized in section VI.

## II. Combined System Description

Fig.1 demonstrates the proposed topology of a combined power generation unit consisting of solar, wind and battery with DC-DC power converters to connect a constant power DC machine as CPL. It can be observed that there are two main branches in the system, thus solar and wind power can compensate each other to a certain degree. The wind power branch includes a wind turbine, a permanent magnet synchronous generator (PMSG), an uncontrolled universal bridge, a single-ended primary inductor converter (SEPIC), wind MPPT controller, and a wind local controller which can close the wind system when the wind speed is too high. The PV power branch is composed by the cascaded boost converters, one MPPT controller based on sliding mode control in the first stage, and a local PI controller in the second stage. Each source runs a maximum power point tracking (MPPT) algorithm and receives the control signal from the supervisory controller. The control system structure and gains are provided in the following sections.

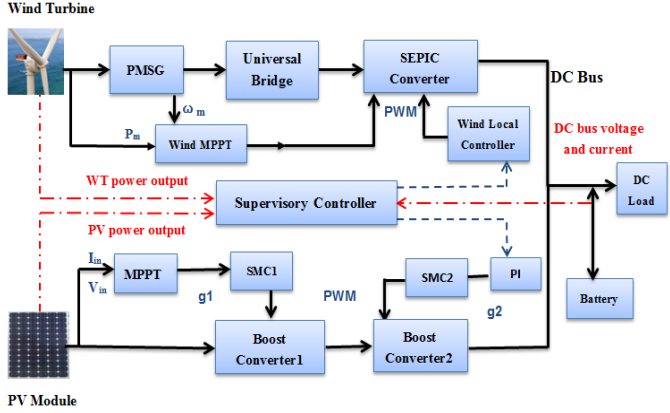

Fig. 1. Configuration of proposed hybrid new energy system

### III. WIND MAXIMUM POWER POINT TRACKING

#### A. Modeling of Wind Power System

Wind system can absorb the wind energy and convert them into electric energy through the power electronics devices. A typical wind generation system is composed of wind turbine, generator, universal bridge, and the power converters. The generator is driven by the torque which comes from the rotation of the wind turbine. Then the universal bridge is used to make the AC power to DC power and supply the source voltage of power converters. The fundamental formula determines the output power of wind turbine is given by:

$$P = \frac{1}{2} p A V^3 C_p \quad , \tag{1}$$

where, ρ is the air density (kg/m3), which is 1.29kg/ m3 in this paper. A is swept area of rotor blads, V is wind speed (m/sec), Cp is the wind power utilization coefficient, the theoretical maximum value of Cp is around 0.593 based on the Betz Theory. The tip speed ratio λ and power conversion value Cp can be obtained as follows:

$$\lambda = \frac{\omega R}{V} = \frac{2\pi R n}{V} \quad , \tag{2}$$

$$C_p(\lambda, \beta) = c_1 \left( \frac{c_2}{\lambda_i} - c_3 \beta - c_4 \right) e^{-\frac{c_5}{\lambda_i}} + c_6 \lambda \quad , \tag{3}$$

$$\frac{1}{\lambda_i} = \frac{1}{\lambda + 0.08\beta} - \frac{0.035}{\beta^3 + 1} \quad . \tag{4}$$

In the above formula, ω is the wind turbine shaft speed (rad/s), n is the wind turbine rotational speed (r/s), $R$ is the radius of the blades ($R=2.5m$ in this paper), and $C_1=0.5176$，$C_2=116$，$C_3=0.4$，$C_4=5$，$C_5=21$，$C_6=0.0068$ [7]. In this paper, the wind power system consists of a fixed pitch angle (0 degrees) and variable speed wind turbine, a perpannent magnet synchronuns generator (PMSG), and a diode bdidge rectifier is built and applied in Matlab/Simulink. The wind turbine torque and rotor speed are equal to the generator torque and rotor speed due to their direct connection without a gearbox. Fig. 2 shows the characteristics of the torque and wind turbine power vs. rotor speed [7].

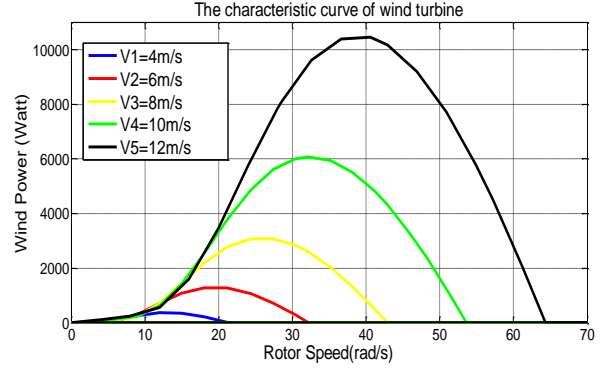

Fig. 2. Wind turbine power vs. rotor speed

#### B. Proposed MPPT Control

As can be seen in Fig. 2 and Fig. 3, there is only one optimal torque and wind mechanical power under different wind speed and every optimal point correspond just one rotor speed. Therefore, the most essential point for tracking the maximum power is to capture the shaft speed of wind turbine at wind speeds. Although the wind MPPT methods are now being implemented through different kinds of power inverters or power converters, these techniques can be categorized into three main types: 1) tip-speed ratio (TSR) control. This method control the optimal shaft speed constant for keeping the best tip-speed ratio and the power utilization coefficient. The advantage of it is the simple implementation. However, because of the requirement of meteorology information, the performance of TSR control method relies on the anemometer accuracy or the estimated wind speed [20]. 2) Power signal feedback (PSF) control. This technology is mainly based on the turbine characteristics curve resulting from simulation or field tests. The data can be stored in form of lookup table and is easy to implement for tracking of the best rotor speed that corresponds the maximum power without wind speed measurement. However, it is difficult to obtain the field data [21]. 3) Perturbation and Observation control. P&O controller generally applies a rotor speed perturbation and observes the output power change to search for the maximum power point. It requires neither characteristics of wind turbine nor the wind speed, so the control method is more flexible and reliable. However, it may lose the effectiveness in large-scale wind turbine systems because of the variation of turbine inertia. In addition, the output of system may have large oscillations in higher power rating systems [22].

In this paper the P&O method is used. From the formula (1) and (2), the wind turbine output power and torque are:

$$P = \frac{1}{2} p A C_p \frac{\omega^3 R^3}{\lambda^3} = K \omega^3, \tag{5}$$

$$T = K \omega^2 \tag{6}$$

where $K=\frac{1}{2}pAC_p\frac{R^3}{\lambda^3}$. When the pitch angle is 0 degree, the $C_p-\lambda$ characteristics is shown in Fig. 3, so the maximum value of $C_{pmax}$ is 0.48 and the optimal $\lambda_{opt}$ is around 8.1.

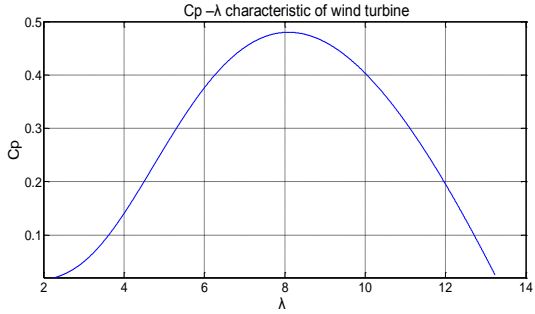

Fig. 3. *Cp –λ characteristic of wind turbine (β=0)*

According to the principle of energy conservation, the input power should equal to the output power, hence the output power decreasing will lead to energy consumed in the generator increased and the rotor speed of generator increased. The output power will increase with the increment of duty cycle of converter and result in the decrement of turbine rotor speed; the output power will decrease with the decrement of duty cycle of converter and result in the increment of turbine rotor speed. Therefore, wind MPPT can be realized through tuning the duty cycle of SEPIC converter to change the turbine shaft speed and keep the power coefficient Cp at maximum. The flow chart of P&O method for wind system is demonstrated in Fig. 4.

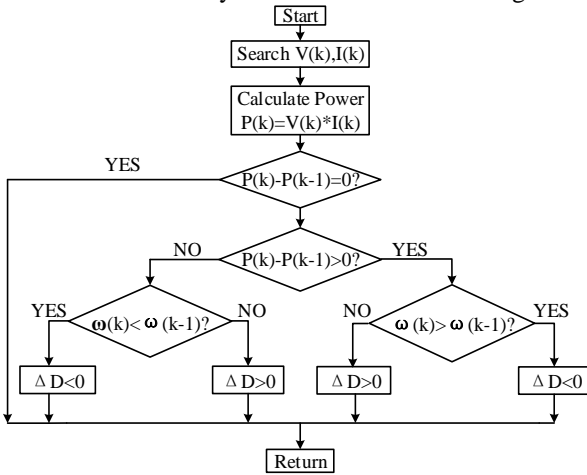

Fig. 4. Flow chart of wind MPPT algorithm.

There are four modes of wind MPPT control in the Fig. 4:
1. $P_k>P_{k-1}$ and $\omega_k>\omega_{k-1}$, means the previous duty cycle perturbation is negative. Thus that the next perturbation become negative;
2. $P_k<P_{k-1}$ and $\omega_k<\omega_{k-1}$, means the previous duty cycle perturbation is positive. Thus the next perturbation to become negative;
3. $P_k>P_{k-1}$ and $\omega_k<\omega_{k-1}$, means the previous duty cycle perturbation is positive. Thus the next perturbation become positive;
4. $P_k<P_{k-1}$ and $\omega_k>\omega_{k-1}$, means the previous duty cycle perturbation is nagative. Thus the next perturbation to become positive.

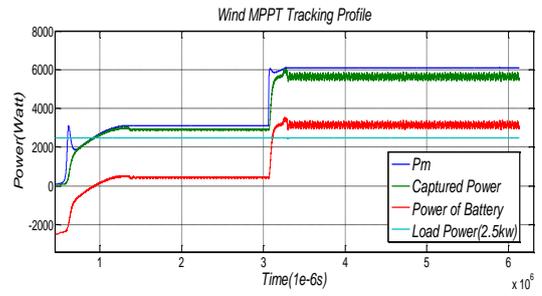

(a) Wind MPPT tracking effect

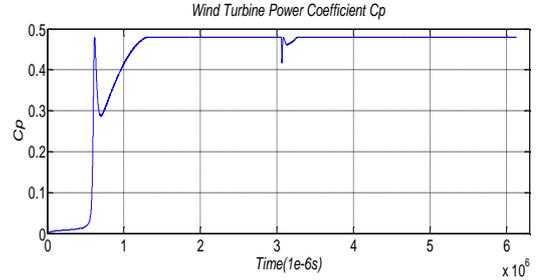

(b) Power conversion coefficient

Fig. 5. Wind MPPT tracking profile to step change in wind speed from 8m/s to 10m/s

Simulations of wind MPPT profile for a DC constant power load (2.5kw) and battery with wind speed variations are illustrated in Figs. 5 and 6. Fig. 5 (a) depicts the MPPT tracking profile and the load power conditions when the wind speed changes from 8m/s to 10m/s. Fig. 5 (b) shows the power coefficient Cp is keeping the maximum value 0.48. Fig. 6 (a) illustrates the MPPT tracking profile and the load power conditions with the variations of wind speed from 12m/s to 10m/s. Fig. 6 (b) also demonstrates the power coefficient Cp reaching the maximum value 0.48.

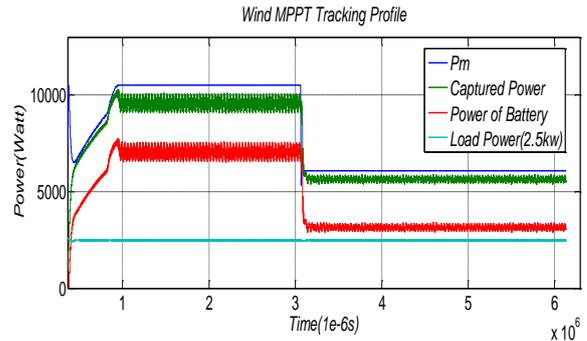

(a) Wind MPPT tracking effect

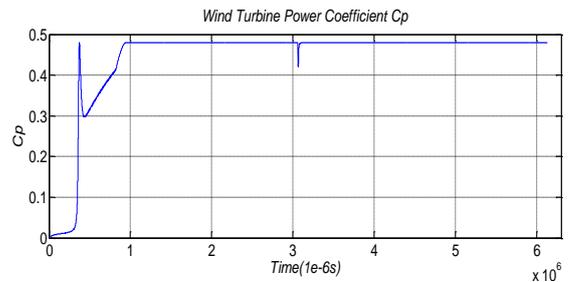

(b) Power conversion coefficient

Fig. 6. Wind MPPT tracking profile to step change in wind speed from 8m/s to 10m/s

## IV. SOLAR MAXIMUM POWER POINT TRACKING

### A. Modeling of Solar Power System

The solar module has been built in Matlab/Simulink based on the equivalent circuit of PV panel in Fig. 7 and the simulation parameters setting in Table I.

TABLE I.
THE SIMULATION PARAMETERS SETTING OF PV PANEL

| $V_{oc}$ | $I_{sc}$ | $V_m$ | $I_m$ | $P_m$ | $R_s$ | $T_{ref}$ | $R_{ref}$ |
|---|---|---|---|---|---|---|---|
| 129V | 19.2A | 105.6V | 17.1A | 1800W | 0.2Ω | 25°C | 1000W/m² |

In the table, $V_{oc}$ is the open circuit voltage of PV panel, $I_{sc}$ is the short circuit current, $V_m$ and $I_m$ are the nominal voltage and currents. $P_m$ in the nominal power under the standard weather climate (25°C and 1000W/m²), $R_s$ is the series resistance of the cell, $T_{ref}$ is the reference temperature and $R_{ref}$ is the radiation at the reference condition.

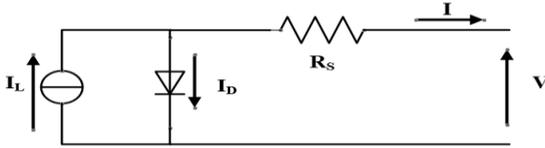

Fig. 7. Equivalent circuit of PV panel

Figure 8 shows the *P-V* curve of the PV model at different solar illumination intensity levels.

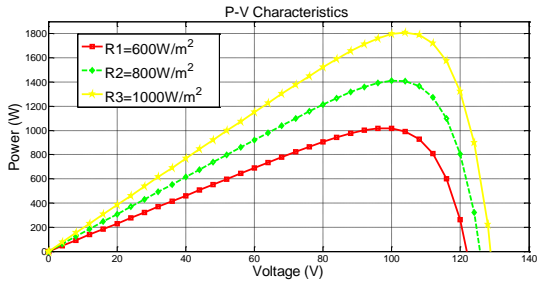

Fig. 8. Power-Voltage characteristic of PV model.

### B. Proposed MPPT Control

To design the MPPT controller according to the sliding mode control theory, it is viewed that the DC power converters have no loss utilizing the concept of ideal canonical elements, such as loss free resistors (LFRs) [23]. The PV system connecting to cascade boost converter and the load is illustrated in Fig. 9 (a). The output power can be represented by the following function of the conductance $g_1=1/r_1$:

$$P_o = g_1 \times V_p^2 \quad . \quad (7)$$

The I–V curve of PV module and the LFR steady-state load-line are also shown in Fig. 9(b). The PV impedance matching is to adjust the intersection of the two curves and obtain the optimal operating point of solar panel by tuning the slope of the load line (conductance g) [24].

The two sliding surfaces for each boost converters are defined as follows:

$$s_1(x) = g_1 v_p - i_{L_1} \quad (8)$$
$$s_2(x) = g_2 v_{c_1} - i_{L_2} \quad (9)$$

where $g_1$ is the conductance of solar module, $v_p$ is output voltage of the PV panel, $i_{L_1}$ is the output current at the first stage, $g_2$ is the constant from PI controller, $v_{c_1}$ is the DC coupling capacitance voltage between the two stages, and $i_{L_2}$ is the current through L$_2$ in second stage [24].

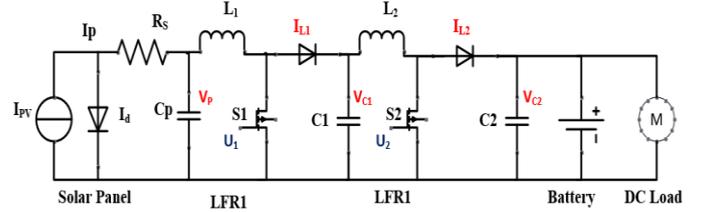

(a)

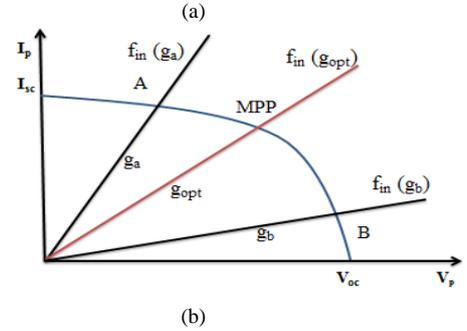

(b)

Fig. 9. (a) PV panel to the load using cascaded boost-based LFRs. (b) PV panel operating points for impedance matching

The control objective of the solar MPPT is to obtain the best operating point and ensure the PV panel operating point around the MPP in spite of the temperature, insulation and load variation. There are several tracking methods applied for different types of DC-DC converters such as fractional open-circuit voltage over short circuit current method, the incremental conductance (INC) method, and the P&O method. Shown in Figure 10, the Extremum Seeking Control (ESC) is used to control the PV system to approach the MPP by increasing or decreasing one suitable control signal in this paper [25,26]. Figure 11 shows the MPP tracking profile with an irradiance change from 1000W/m² to 800W/m².

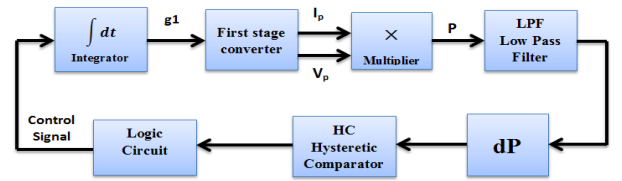

Fig. 10. PV MPPT scheme based on ESC [25].

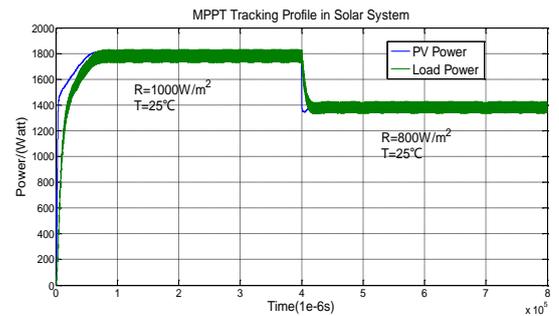

Fig. 11. Solar MPPT tracking profile.

## V. SUPERVISORY CONTROLLER

In a DC power system where multiple power converters supply energy to a common bus, the current sharing and system protection are essential. Therefore, the supervisory controller decides the type of local sources according to the four system operating contingencies (Table II) [27], [28]. The net power that battery needs to provide $\Delta P$ can be obtained from the renewable energy and load power deficit as follows:

$$\Delta P = P_s + P_w - P_{load} \quad (10)$$

where $P_s$ is the solar power, $P_w$ is the wind power, $P_{load}$ is the power demand from the constant power load.

TABLE II.
SUPERVISORY CONTROL CONTINGENCIES

| Mode | Condition | Control Effect |
|---|---|---|
| 1 | $\Delta P \geq 0$, $0 \leq SOC \leq 95\%$ | Feeding Load and charge Battery |
| 2 | $\Delta P \geq 0$, $SOC > 95\%$ | Feeding Load and battery (Surplus power through a dummy load) |
| 3 | $\Delta P \leq 0$, $SOC < 40\%$ | Charge battery and off load |
| 4 | $\Delta P \leq 0$, $SOC > 40\%$ | Feeding load and battery discharge to the load |

The simulation results for the dc constant load and battery that show the supervisory control effect for the combined power system are depicted in Fig. 12-14 respectively. Fig. 12 mainly demonstrates the mode 1 and mode 4, where the load power reference value at 5kw and the state of charge (SOC) of a 900Ah battery is 50%. It shows the MPP tracking effects of the solar system and wind system, and the response of the battery power with a DC load power during the irradiance change from 600W/m² to 900W/m² of PV and 8m/s to 9m/s of wind energy system. The DC micro-grid is capable of charging and discharging the battery at the required rate automatically while capturing the maximum power from the solar panel and wind system under variations of weather condition.

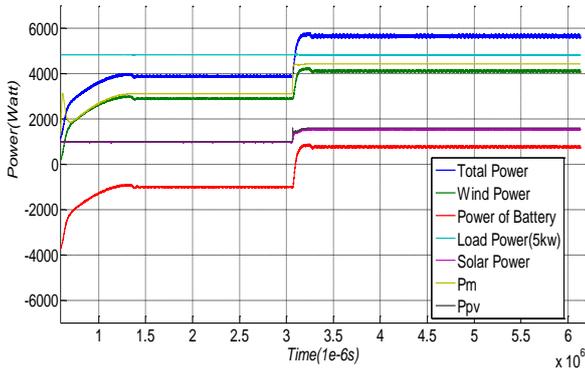

Fig. 12 MPPT Tracking profile for combined power generation unit under a step change in solar irradiance from 600W/m² to 900W/m² and a step change in wind speed from 8m/s to 9m/s.

Fig. 13 illustrates the performance of control in mode 2, where the load power reduces to 3kw and the SOC of battery is 98%. In this case, the solar irradiance experienced a change from 1000 W/m² to 700 W/m² and the wind speed changed from 7m/s to 8m/s. Now the battery is still be floating charged to maintain the capacity in case of self-discharge.

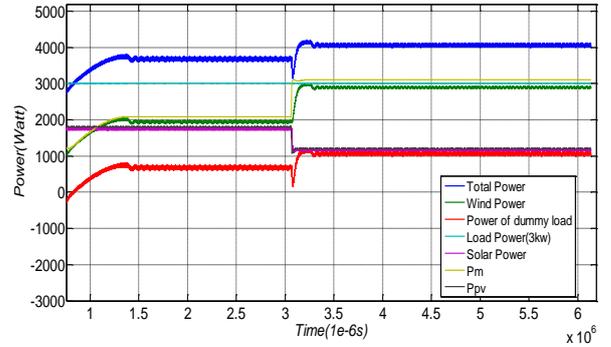

Fig. 13. Mode 2 of supervisory control strategy

Fig. 14 depicts the performance of control in mode 3, where the load power is 5kw and the SOC of battery is 30%. In this case, the solar irradiance changed from 800 W/m² to 600 W/m² and the wind speed changed from 8m/s to 7m/s, which results in $\Delta P$ was not larger than zero and SOC lower than 40%. Therefore, the load should be off and the total power captured from renewable sources was used to charge the battery without satisfying 5kW DC load power.

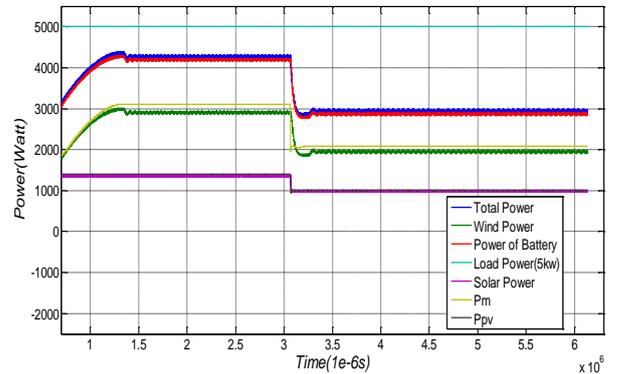

Fig. 14. Mode 3 of the supervisory control strategy

## VI. CONCLUSION

This paper presented the modeling and control of the wind-solar-storage power generation unit. Models of the horizontal axis variable speed wind turbine and a PV array, power converters, their MPPT controllers were established in Matlab/Simulink. In wind power system, the P&O method was utilized and the MPPT controllers based on the sliding mode control theory was applied. Both methods illustrated a stable and effective tracking performance with variations of wind speed and solar irradiation under constant power load. The supervisory control strategy which has four conditions was also proposed to capture the maximum power from each renewable energy sources while connected to a common dc bus. A robust and smooth control effect was obtained in both power sources. Simulation results demonstrated an accurate operation and functionality of the presented method.